\documentclass[paper,notoc,nohyper]{JHEP}
\usepackage{amsmath}
\usepackage{mathrsfs}
\usepackage{epsfig}
\newcommand{\lf}[2] {\mbox{$\frac{#1}{#2}$}}
\def\epem{e$^+$e$^-$ }
\def\as{\ensuremath{\alpha_{s}}}
\def\rd{\mathrm{d}}
\def\bs{\mathbf}
\renewcommand\O[1]{{\cal O}(#1)}

\def\Nf{N{\!_f}}
\def\Bmin{B_N}
\def\Bmax{B_W}
\def\Light{\rho_L}
\def\Heavy{\rho_H}

\title{\boldmath Resumming the Light Hemisphere Mass  and Narrow Jet Broadening 
distributions \\in e$^{+}$e$^{-}$
  annihilation\footnote{Work supported in part by the UK Particle Physics and
Astronomy Research Council and by the EU Fourth Framework Programme
`Training and Mobility of Researchers', Network `Quantum Chromodynamics
and the Deep Structure of Elementary Particles',
contract FMRX-CT98-0194 (DG 12 - MIHT).
}}
\author{
S.~J.~Burby$^a$ and E.~W.~N.~Glover$^b$\\
$^a$Department of Theoretical Physics, Lund University,\\
 S\"olvegatan 14A, S-223 62
 Lund, Sweden\\[1mm]
$^b$Department of Physics, 
University of Durham, 
Durham DH1 3LE, 
England\\[1mm]
E-mail: \email{Stephen@thep.lu.se}, \email{E.W.N.Glover@durham.ac.uk}}
\abstract{
We present predictions of two event shape distributions, the light
  hemisphere mass and the narrow jet broadening, to next-to-leading
  logarithmic order.  We apply the coherent branching formalism to resum the
  leading $\O{\as^n L^{2n-1}}$ and next-to-leading $\O{\as^n L^{2n-2}}$ 
  infrared logarithms to all orders in the coupling constant.  We include
  the recently calculated non-logarithmic next-to-leading order
  contributions.  Applying simple power corrections to the resummed result
  gives good agreement with the available data from LEP.
}
\keywords{QCD, Jets, LEP HERA and SLC Physics, NLO and NNLO Computations}
\preprint{{LUTP/01/03},{~DCTP/01/06},{~IPPP/01/03},{~hep-ph/0101226}}

\begin{document}

\section{Introduction}

Over the last decade, event shape distributions in \epem annihilation have
provided some of the most precise tests of QCD.  Most of the attention has
focussed on three-jet event shape variables such as thrust, wide jet
broadening, heavy hemisphere mass, C parameter etc..  These variables may be
generated by single gluon emission from the underlying $q \bar q$ pair in the
\epem annihilation process which first contributes at $\O{\as}$.  The analysis
of such events is rather sophisticated. In the first instance, fixed order QCD
perturbation theory \cite{ERT} is employed (at next-to-leading order (NLO)).
However, when the observable $O$ is small and $L = \ln(1/O)$ is large the
distribution suffers large logarithmic contributions of the form $\O{\as^n
L^{2n-1}}$  that render the fixed order perturbation theory insufficient.    In
many cases the dominant infrared logarithms can be resummed using the coherent
branching formalism \cite{coherent,mass,heavy,wide}. Furthermore, significant
power suppressed effects are present and have been phenomenologically studied
\cite{power}.  To accurately describe the data,  all of these contributions --
fixed order, infrared resummation and power correction --  are found to be
necessary.  Perhaps the most important measurement extracted from three-jet
event shape data is the determination of the strong coupling constant.

Four-jet event shape observables also contain useful information about QCD.
They are more sensitive to the triple gluon vertex and therefore the true gauge
structure of QCD \cite{gauge} and also to the presence of other light coloured
particles such as the gluino \cite{gluino} that can be pair produced by gluon
splitting. However, these variables have received much less attention partly
because they are suppressed by an additional power of $\as$ requiring a second
gluon to be radiated but also because the theoretical description is much less
developed.   Recently however,  four separate general purpose Monte Carlo
programs have been  developed to estimate the next-to-leading order,
$\O{\as^3}$, corrections to four-jet event shapes: {\tt MENLO PARC}
\cite{menloparc}, {\tt DEBRECEN} \cite{debrecen} and {\tt MERCUTIO}
\cite{mercutio} employing on the one-loop helicity amplitudes for  $e^+e^- \to
4$~partons \cite{BDKW} and {\tt EERAD2} \cite{eerad2} based on the interference
of the one-loop matrix element with tree level \cite{CGM}.  For most four-jet
event shape variables, the situation is the same as for the three-jet event
shape variables; the NLO corrections are very large and, for renormalisation
scales of the order of the centre-of-mass energy $Q$, still undershoot the data
significantly \cite{eerad2}.   This points at the presence of large infrared
effects as well as significant power corrections.  With the exception of the
four-jet rate in certain schemes \cite{4jet} and near-to-planar three-jet
events \cite{BDMZ}, the issue of resumming infrared logarithms for specific
four-jet event shape variables has not yet been addressed.  Similarly, power
suppressed effects have only been studied for the out-of-event plane momentum
distribution \cite{BDMZ}. It is the goal of this Letter to resum the leading
$\O{\as^n L^{2n-1}}$ and next-to-leading $\O{\as^n L^{2n-2}}$ infrared
logarithms  to all orders  in the coupling constant for the light hemisphere
mass $\Light$ and narrow jet broadening $\Bmin$ distributions.

To be more precise let us first define these variables.   We first 
separate the event at centre-of-mass energy $Q = \sqrt{s}$ into two hemispheres $H_1$, $H_2$ divided by the plane
normal to the thrust axis ${\bf n}_T$.  Particles that satisfy
${\bf p}_i.{\bf n}_T > 0$ are assigned to hemisphere $H_1$, while all other
particles are in $H_2$. Jet broadening measures the summed scalar momentum
transverse to the thrust axis in one of the hemispheres 
while the hemisphere mass is
the invariant mass of the hemisphere,
\begin{eqnarray}
\Bmin &=&\min_{i=1,2} \frac{\sum_{{\bf p}_k \in H_i} 
 |{\bf p}_k \times {\bf n}_T|}{2\sum_k|{\bf p}_k|}\\
\Light &=& \frac{1}{s} \cdot \min_{i=1,2}
 \left( \sum_{{\bf p}_k \in H_i} p_k \right)^2.
\end{eqnarray}
Note that this definition of the light hemisphere mass is the common 
modification of the original variables suggested by Clavelli \cite{clavelli}.
These four-jet event shape variables are intimately connected to 
their three-jet event shape counterparts, the wide
jet broadening and heavy hemisphere mass,
\begin{eqnarray}
\Bmax &=&\max_{i=1,2} \frac{\sum_{{\bf p}_k \in H_i} 
 |{\bf p}_k \times {\bf n}_T|}{2\sum_k|{\bf p}_k|}\\
\Heavy &=& \frac{1}{s} \cdot \max_{i=1,2}
 \left( \sum_{{\bf p}_k \in H_i} p_k \right)^2
\end{eqnarray}
that have the property of exponentiation. That is to say that 
 the fraction of events where
the observable $O = \Bmax$ or $O = \Heavy$ 
has a value less than $O$ obeys the exponentiated form,
\cite{heavy,wide}
\begin{eqnarray}
R(O,\as(Q^2)) &=& \int_0^O \frac{1}{\sigma} \frac{d\sigma}{dO} dO   \\
\label{eq:expo}
&=& C(\as(Q^2)) \Sigma(O,\as(Q^2)) + D(O,\as(Q^2))
\end{eqnarray}
where 
\begin{eqnarray}
C(\as(Q^2)) &=& 1 + \sum_{n=1}^\infty  C_n \as^n  \\
\ln\biggl(\Sigma(O,\as(Q^2))\biggr) &=& \sum_{n=1}^\infty \sum_{m=1}^{n+1} G_{nm} \as^n L^m
\nonumber \\
&=& Lg_1(\as L) + g_2(\as L) + \as g_3(\as L) + \ldots   \\
D(O,\as(Q^2)) &=& \sum_{n=0}^\infty D_n \as^n.
\end{eqnarray}
Here $C_n$ and $G_{nm}$ are constants and the perturbatively calculable
coefficients $D_n \to 0$ as $O \to 0$. Knowledge of 
$G_{n n+1}$ (or equivalently $g_1$) together with $G_{11}$ 
allows resummation of terms in $\Sigma$ down to $\O{\as^n L^{2n-1}}$.
Partial inclusion of other subleading logarithms may be accomplished by
retaining all knowledge of $g_2$.  
The accuracy of $R$ also depends on the number of terms known in $C(\as(Q^2)$, 
with
each
known term giving information about two towers of logarithms.  For example, 
$C=1$ resums the $\O{\as^n L^{2n}}$ and $\O{\as^n L^{2n-1}}$ terms in $R$, 
$C_1$ resums the $\O{\as^n L^{2n-2}}$ and $\O{\as^n L^{2n-3}}$ terms and so on.

\section{Coherent branching}

The coherent branching formalism allows the resummation of soft and collinear
logarithms due to the emission of gluons from a hard parton.   As a specific
example, let us consider the  
jet mass distribution $J^a(Q,k^2)$ as the probability of
producing a final state jet with invariant mass $k^2$ from a parent parton
$a$ produced in a hard process at the scale $Q^2$. For an initial quark
this is \cite{mass}
\begin{eqnarray}
\label{eq:cbranch}
\lefteqn{  J^q(Q^2,k^2)=\delta(k^2) + \int\limits^{Q^2}_0 \frac{\rd
  \tilde{q}^2}{\tilde{q}^2} \int\limits^1_0 \rd z\
  \Theta(z^2(1-z)^2\tilde{q}^2-Q^2_0) 
P^{qq}[\as(z^2(1-z)^2)\tilde{q}^2),z]}\nonumber\\
&\times& \Biggl[\int\limits^{\infty}_0 \rd q^2 \int\limits^{\infty}_0 \rd {k'}^2
  \delta\left(k^2-z(1-z)\tilde{q}^2-\frac{{k'}^2}{z} -
  \frac{q^2}{1-z}\right) J^q(z^2\tilde{q}^2,{k'}^2) 
J^g((1-z)^2\tilde{q}^2,q^2)\nonumber\\
&&-J^q(\tilde{q}^2,k^2)\Biggr]
\end{eqnarray}
normalised such that
\begin{equation}
    \int^\infty_0 dk^2 
    J^a(Q^2,k^2)= 1
\end{equation}
and where the next-to-leading order $q \to q g$ splitting kernel in the $\overline{MS}$
scheme with $\Nf$ flavours is
\begin{equation}
P_{qq}\left[\as,z\right] = \frac{\as}{2\pi} C_F \frac{1+z^2}{1-z}
\left(1 +\frac{\as}{2\pi} K \right)
 + \ldots,
\end{equation}
where
\begin{equation}
\label{eq:K}
K =  C_A \left(\frac{67}{18} - \frac{\pi^2}{6}\right) 
    - \frac{5}{9} \Nf .
\end{equation}   
Eq.~(\ref{eq:cbranch}) has a simple physical interpretation.   The first term is
the possibility that the originating parton does not emit any radiation.  The
transverse momentum of the parton is therefore unchanged.   Alternatively, the 
quark may branch into a quark and a gluon subject to the phase space constraints of two body 
decay and which subsequently undergo further emissions.   This is described by
the term proportional to $J^q ~J^g$.   The last term is due to virtual
corrections and ensures that soft and collinear singularities are regularised.
A similar equation holds for $J^g$ and involves the $g \to gg$ and $g \to q \bar q$ 
splitting kernels.  Solving the integral equation for
$J^a$ is equivalent to resumming the infrared logarithms.
In particular, the probability that an isolated parton $a$ forms a jet with mass less than
$\rho Q^2$ is given by
\begin{equation}
\label{eq:normal}
\Sigma^a(\rho,\as(Q^2)) = \int^{\rho Q^2}_0 dk^2 J^a(Q^2,k^2).
\end{equation} 
In Ref.~\cite{mass}, $\ln(\Sigma^a_H)$ is solved to next-to-leading 
logarithmic accuracy.

Similarly we can define the function $T^a(Q,{\bf k}_t;p_t)$ \cite{wide} which
describes the distribution of the summed scalar transverse momentum $p_t$ in a jet
of parton type $a$ produced with vector transverse momentum ${\bf k}_t$, at
scale Q. The structure is identical to Eq.~(\ref{eq:cbranch}).

\begin{figure}[t]
\begin{center}
~\epsfig{file=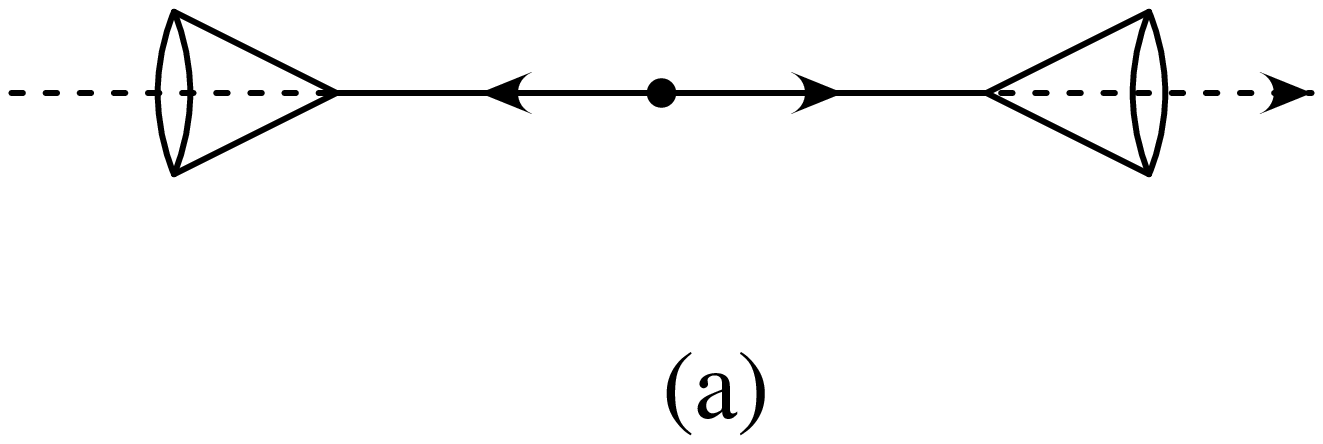,width=6.5cm}
\hspace{0.5cm}
~\epsfig{file=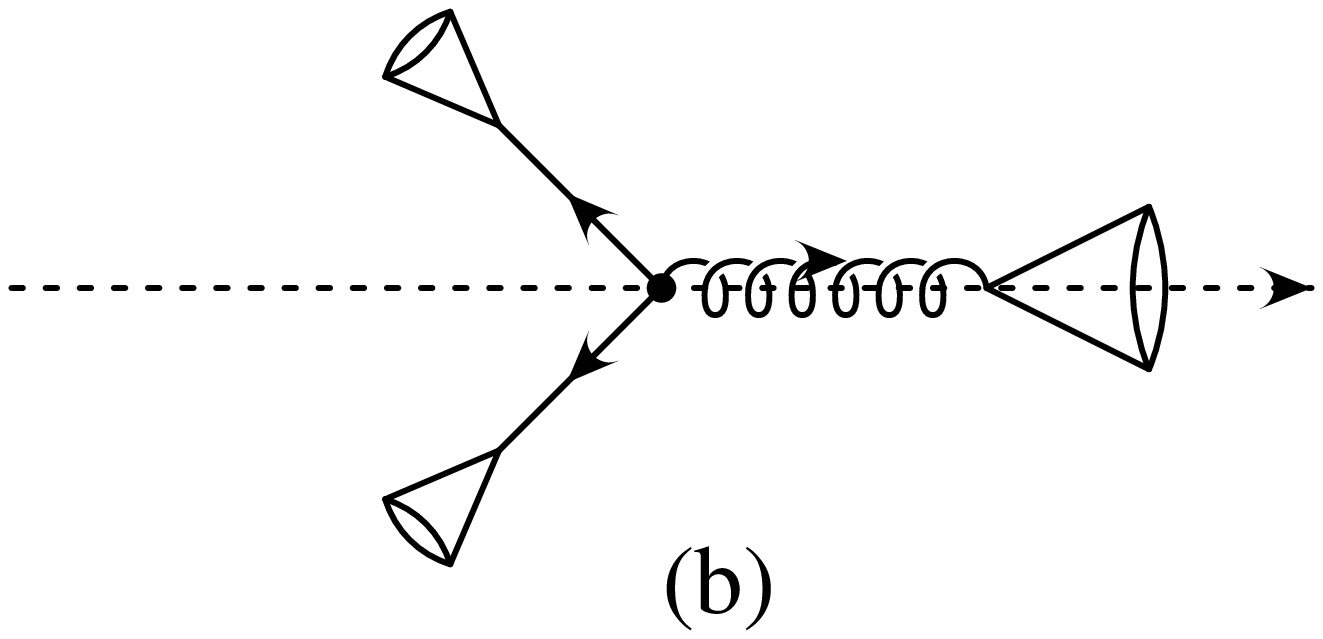,width=6.5cm}
\end{center}
\caption[]{Event pictures of (a) two-jet configurations from quark-antiquark 
final states and (b) three-jet configurations originating from $q \bar q g$
events where the gluon is the hardest parton.  
The cones represent coherent soft
and collinear gluon emission. The thrust axis is denoted by a dashed line.}
\label{fig:pic}
\end{figure}

\section{The probabilistic interpretation}

We can apply the coherent branching formalism to event shapes in the following
schematic way illustrated in Fig.~1. The underlying configuration in \epem
annihilation is the production of a quark-antiquark pair aligned with the
thrust axis (Fig.~1(a)).
Each parton then undergoes soft and collinear gluon
emission (denoted by the open cone at the head of the parton).   This
contribution describes small angle and soft emission accurately in 
{\em two-jet-like}
events  when $\Bmax$ and $\Heavy$ are small and gives rise to the exponentiated
first term in Eq.~(\ref{eq:expo}). However, it does not describe the possibility
of wide angle gluon emission shown in Fig.~1(b) 
where the gluon is the hardest parton.  This  {\em three-jet-like} 
matrix element correction is not logarithmically enhanced at
$\O{\as}$ and first contributes at next-to-next-to-leading logarithmic accuracy
and can therefore be neglected.

Let us first focus on the hemisphere masses.  To next-to-leading logarithmic
accuracy, the fraction of events with
heavy hemisphere mass less than $\rho_H$ i.e. $k_1^2 < \rho_H Q^2$ and $k_2^2
< \rho_H Q^2$ is given by the two jet contribution 
\begin{eqnarray}
\lefteqn{R_H(\rho_H,\as(Q^2))\ \mathop{=}_{\rho_H\ll1} }\nonumber \\ 
&& \int\limits^{\infty}_0 \rd k_1^2
 \int\limits^{\infty}_0 \rd k_2^2\ J^q(Q^2,k_1^2)\ J^q(Q^2,k_2^2)\
 \Theta(\rho_H Q^2-k_1^2)\ \Theta(k_1^2-k_2^2) \nonumber \\ 
&&+ \int\limits^{\infty}_0 \rd k_1^2
 \int\limits^{\infty}_0 \rd k_2^2\ J^q(Q^2,k_1^2)\ J^q(Q^2,k_2^2)\
 \Theta(\rho_H Q^2-k_2^2)\ \Theta(k_2^2-k_1^2) ,
\end{eqnarray}
and where the constraint that $\rho_H\ll1$ has suppressed the three and more jet
contributions.
Following the steps of Ref.~\cite{wide}, we can rewrite this formula using the
phase space restrictions as
\begin{equation}
R_H(\rho_H,\as(Q^2))\ \mathop{=}_{\rho_H\ll1} 
\left [ \Sigma^q_H(\rho_H,\as(Q^2)) \right]^2.
\end{equation}
where
\begin{equation}
\label{eq:sigmaqH}
\Sigma^q_H(\rho_H,\as(Q^2)) = \int\limits^{\rho_H Q^2}_0 \rd k^2\ J^q(Q^2,k^2).
\end{equation}

The fraction of events with light hemisphere mass less
than $\rho_L$ also receives contributions
from the two-jet configuration
when $k_1^2 < \rho_L Q^2$ and $k_2^2 > k_1^2$ and vice versa.  
Altogether we have,
\begin{eqnarray}
\lefteqn{R_L(\rho_L,\as(Q^2))\ \mathop{=}_{\rho_L\ll1} }  \nonumber  \\
&&  \int\limits^{\infty}_0 \rd k_1^2
 \int\limits^{\infty}_0 \rd k_2^2\ J^q(Q^2,k_1^2)\ J^q(Q^2,k_2^2)\
 \Theta(\rho_L Q^2-k_1^2)\ \Theta(k_2^2-k_1^2)  \nonumber\\
& & +\int\limits^{\infty}_0 \rd k_1^2
 \int\limits^{\infty}_0 \rd k_2^2\ J^q(Q^2,k_1^2)\ J^q(Q^2,k_2^2)\
 \Theta(\rho_L Q^2-k_2^2)\ \Theta(k_1^2-k_2^2).
\end{eqnarray}
Simplifying the phase space constraints,
 we find
\begin{equation}
R_L(\rho_L,\as(Q^2))\ 
\label{eq:rl}
\mathop{=}_{\rho_L\ll 1} 2~\Sigma^q_H(\rho_L,\as(Q^2)) 
- \left[\Sigma^q_H(\rho_L,\as(Q^2))\right]^2.  
\end{equation}
The functions that resum the logarithms for the light hemisphere mass are the
same as those that resum the logarithms for the heavy hemisphere mass. Now
however, exponentiation in its purest form is spoiled because the final result
is a sum of terms.

The analysis for the narrow jet broadening proceeds in the same way.  We
find that $R_N$, the probability of finding an event with a light hemisphere
mass less than $B_N$, is given by
\begin{equation}
R_N(B_N,\as(Q^2))\ 
\label{eq:rn}
\mathop{=}_{B_N\ll1} 2~\Sigma^q_W(B_N,\as(Q^2)) 
- \left[\Sigma^q_W(B_N,\as(Q^2))\right]^2 
\end{equation}
where the probability of obtaining a jet with summed scalar 
transverse momentum $p_t$ with
respect to the jet axis less than
$2 B Q$ starting from a parton of type $a$, $\Sigma^a_W$, is given by,
\begin{equation}
\Sigma^a_W(B,\as(Q^2) = \int_0^{2BQ} T^a(Q,\bs{0},p_t) dp_t. 
\end{equation}

\section{All-orders resummation of large logarithms}

In this section we discuss the all-orders resummation of leading logarithms
$\O{\as^n L^{2n}}$ and next-to-leading $\O{\as^n L^{2n-1}}$ logarithms to all
orders in the coupling constant. From eqs.~(\ref{eq:rl}) and (\ref{eq:rn}),
we see that to determine $R_L$ and $R_N$ requires knowledge of $\Sigma^a_H$
and $\Sigma^a_W$ respectively.  Both of these functions have the
exponentiated form of Eq.~(\ref{eq:expo}) and the corresponding functions $g_1(\as
L)$ and $g_2(\as L)$ determined.

Explicit expressions for $\Sigma_H$ valid to this order are given in
\cite{mass} and, introducing the renormalisation scale dependence in the
standard manner and dropping
the parton index, we reproduce them here for illustrative purposes,
\begin{eqnarray}
\label{eq:sigH}
\Sigma_H(\rho,\as(\mu^2),\lf{Q^2}{\mu^2}) &=& 
\frac{\exp[\mathcal{F}(\as(\mu^2),L)]}
{\Gamma[1-\mathcal{S}(\as(\mu^2),L)]}\nonumber \\
&=& \frac{\exp[L f_1(x)+f_2(x) + x^2 f_1^\prime(x)\ln(\mu^2/Q^2)]}
{\Gamma[1-f_1(x)-x f_1^\prime(x)]} + \O{\as^n L^{n-1}}
\end{eqnarray}
where
\begin{equation}
L = \ln(1/\rho), \qquad x = \beta_0 \as(\mu^2) L.
\end{equation}
The functions $f_1$, $f_2$ and $f_1^\prime$ are 
\begin{eqnarray}
f_1(x) &=& -\frac{A^{(1)}}{2\pi\beta_0 x}
\Biggl [ 
(1-2x)\ln(1-2x)-2(1-x)\ln(1-x)
\Biggr], \\
f_2(x) &=& -\frac{A^{(2)}}{2\pi^2\beta_0^2}
\Biggl [ 2\ln(1-x)-\ln(1-2x)\Biggr ]\nonumber \\
&&
+\frac{B^{(1)}}{2\pi\beta_0} \ln(1-x)
-\frac{A^{(1)}\gamma_E}{\pi\beta_0}
\Biggl [ \ln(1-x)-\ln(1-2x)\Biggr ]\nonumber \\
&&
-\frac{A^{(1)}\beta_1}{2\pi\beta_0^3}
\Biggl [ \ln(1-2x)-2\ln(1-x)+\frac{1}{2}\ln^2(1-2x)-\ln^2(1-x)\Biggr ],\nonumber \\
 \\
f_1^\prime(x)&=&\frac{A^{(1)}}{2\pi\beta_0x^2}
\Biggl [ \ln(1-2x)-2\ln(1-x)\Biggr ], 
\end{eqnarray}
with
\begin{equation}
\beta_0 = \frac{11C_A-2\Nf}{12\pi},\qquad \beta_1 =
\frac{17{C_A}^2-5C_A\Nf-3C_F\Nf}{24\pi^2},
\end{equation}
For quarks, 
\begin{equation}
\label{eq:A1q}
A^{(1)} = C_F, \qquad A^{(2)} = \frac{1}{2}C_F K, 
\qquad B^{(1)} = -\frac{3}{2}C_F,
\end{equation}
with $K$ given by Eq.~(\ref{eq:K}).

Altogether Eqs.~(\ref{eq:sigH}) to (\ref{eq:A1q}) are sufficient to determine 
$R_N$ of Eq.~(\ref{eq:rn}) to next-to-leading logarithmic accuracy.
The expression for $R_N$ therefore correctly sums to all orders in
the strong coupling, \as, only the leading two towers of large logarithms
from $\O{\as^n L^{2n}}$ down to $\O{\as^n L^{2n-1}}$.

\begin{figure}[t]
  \begin{center}
    \vspace{1cm}
    \epsfig{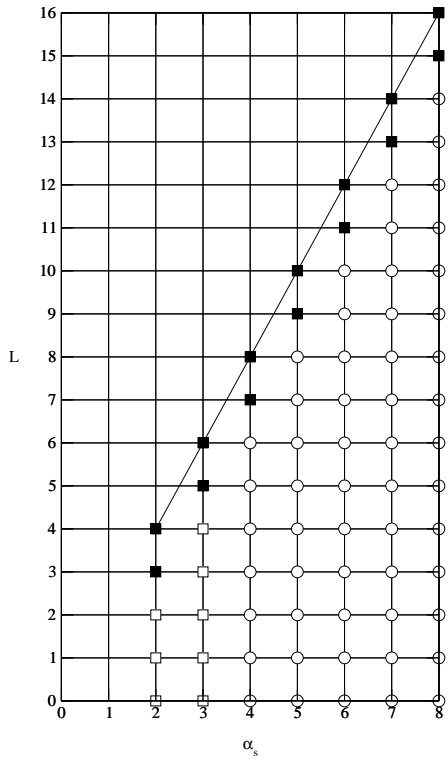}
    \caption[Towers of infrared logarithms in four-jet event shape rates]
    {The towers of infrared logarithms appearing in the four-jet event shape
      rates $R_L$ and $R_N$. The resummation includes the leading and
      next-to-leading logarithms denoted by filled squares and 
      the complete $\O{\as^2,\as^3}$
      contributions from fixed order denoted by the empty squares. All other
      terms are incomplete and denoted by empty circles.  The black filled
      squares denote terms generated purely in the two-jet limit.}
    \label{fig:towers}
  \end{center}
\end{figure}

The resummed formula (\ref{eq:rl}) does not include the $\O{\as^2 L^2}$ or
$\O{\as^2 L}$ terms (just as the analogous formulae for resumming three jet
variables do not include the $\O{\as^2 L^2}$ or $\O{\as^2 L}$ terms) present in
the lowest order perturbative coefficient. Similarly, it does not produce the
$\O{\as^3 L^4}$ to $\O{\as^3 L}$ terms that occur in the next-to-leading order
perturbative coefficient. The perturbative calculation provides the $\as^2$ and
$\as^3$ contributions exactly and therefore the most significant omitted term
is $\O{\as^4 L^6}$, see Fig.~\ref{fig:towers}.

Precisely the same discussion applies to the narrow jet broadening.   Using the
coherent branching formalism the functions  $g_1(\as L)$ and $g_2(\as L)$ have 
been determined for $\Sigma_W$   and Catani et al (CTW)  have provided
analogous expressions for $\Sigma_W$ that are given in \cite{wide}.   However, 
in doing so certain simplifying approximations concerning the  recoil
transverse momentum have been made.      Dokshitzer and collaborators
\cite{recoil} have found that treating the quark recoil more carefully causes
the CTW result for $\Sigma_W$ to be adjusted by a multiplicative factor.  This
modifies $g_2(\as L)$ but leaves $G_{11}$ unchanged.  The same correction is in
principle needed for the narrow jet broadening but is relevant beyond the
next-to-leading logarithmic accuracy. Nevertheless, we chose  to apply the 
correction to the CTW result. We do not give the final form for $\Sigma_W$
here, but instead refer the interested reader to Ref.~\cite{recoil}. Inserting
the recoil-corrected form for $\Sigma_W$ in the resummed expression for $R_N$
(\ref{eq:rn}) again allows resummation of the leading and next-to-leading
logarithms as shown in Fig.~\ref{fig:towers}.

\section{Numerical results}

As usual, the resummed result contains part of the fixed order perturbative
contribution and the overlap must be removed by matching. This is done by
expanding the resummed result as a series in the strong coupling constant and
explicitly removing the terms corresponding to the fixed order calculations.
This can be achieved in several ways of which $R$ matching and $\ln(R)$
matching are the most common.  In the $R$ matching scheme, the coefficients
of each of the unsummed logarithms present in the fixed order perturbative
coefficients must be numerically extracted. For the four-jet event shape
observables discussed here, this corresponds to determining the coefficient
of $\as^2 L$ from the lowest order perturbative contribution and the
coefficients of $\as^3 L^2$ and $\as^3 L$ from the next-to-leading order
contribution. This is impractical.  However, in the more commonly used
$\ln(R)$ matching scheme, it is assumed that the fixed order result
exponentiates and therefore it is not necessary to make this extraction
because any logarithmic terms remaining after subtracting the overlap from
the fixed order contribution are exponentially suppressed. 
Of course, for the four-jet variables considered here, $R$ does not
exponentiate.   Nevertheless, in the $\ln(R)$ matching procedure, any
remaining logarithmic terms are still exponentially suppressed. 
We therefore
employ the $\ln(R)$ matching procedure.

To be more precise, the resummed 
result has the form
\begin{equation}
R_{resummed} = \mathcal{C}_1 2 \Sigma - \mathcal{C}_2 \Sigma^2 + \O{\as},
\end{equation}
where $\mathcal{C}_i$ are power series in $\as$ and $\O{\as}$ represents the
contribution from three-jet configurations.  The $\O{\as}$ term of
$\mathcal{C}_1$ and $\mathcal{C}_2$ are related by the constraint that there
can be no logarithmic contributions at $\O{\as}$ - the expansion of $R$
containing logarithms starts at $\O{\as^2}$.  However, to the
next-to-leading logarithmic accuracy relevant for this paper, we can set
$\mathcal{C}_1 = \mathcal{C}_2 = 1$.  The $\ln{R}$ matching formula then
dictates that
\begin{equation}
\label{eq:lnR}
\ln(R) = \ln(R_{resummed}) + \ln(R_{fixed}) - \ln(R_{expanded}),
\end{equation}
where $\ln{R_{fixed}}$ and $\ln{R_{expanded}}$ are obtained by expanding the
fixed order result and the $\ln{R_{resummed}}$ through to $\O{\as^3}$
respectively.  Using the expansion of $\ln(\Sigma)$ as a series,
$$
\ln(\Sigma) = \sum_{n=1}^{\infty} \sum_{m=1}^{n+1} G_{nm} \as^n L^m,
$$
and with $\mathcal{C}_1 = \mathcal{C}_2 = 1$, we see that,
\begin{eqnarray}
\ln(R_{expanded}) &=& -(G_{11}L+G_{12}L^2)^2 \as^2\nonumber \\
&&-2(G_{11}L+G_{12}L^2)(G_{21}L+G_{22}L^2+G_{23}L^3+\frac{1}{2}(G_{11} L
+G_{12}L^2)^2 )\as^3 \nonumber\\
&&+ \O{\as^4}.
\end{eqnarray}
At next-to-leading logarithmic accuracy, logarithmic terms from the fixed order
contribution remain in $\ln(R)$
given by Eq.~(\ref{eq:lnR}), however, they are exponentially suppressed when 
translated back to $R$.

We expect that at large values of the observable $O$ the resummed result is
dominated by the fixed order calculation. However, the resummed expressions
(\ref{eq:rl}) and (\ref{eq:rn}) valid in the small $\Light$ and $\Bmin$ limits
do not contain information about the kinematic endpoints of the distributions. 
To ensure that the resummed result vanishes at the 
endpoint we make the substitution
\begin{equation}
\label{eq:mod}
\frac{1}{O} \to \frac{1}{O}-\frac{1}{O^{\rm max}} +1,
\end{equation}
where $O^{\rm max}$ corresponds to the endpoint of the distribution at the
accuracy of the fixed order calculation.
We use
$\Light^{\rm max} = 0.167$ and $\Bmin^{\rm max} = 0.204$.

\begin{figure}
\begin{center}
~\epsfig{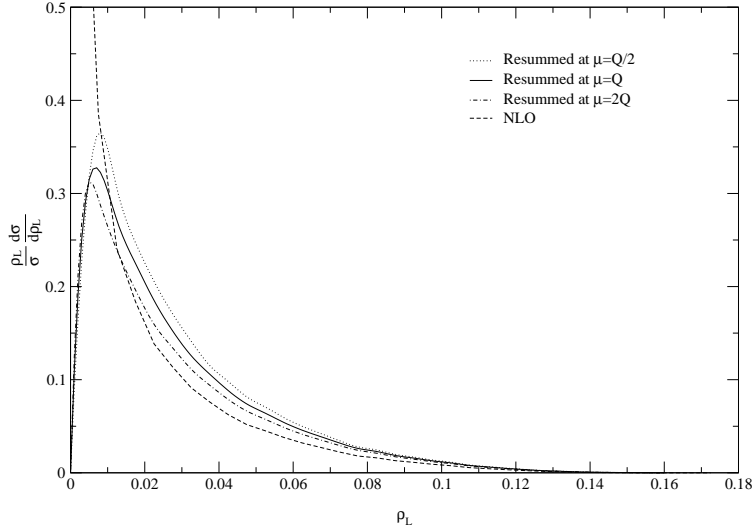}
\end{center}
\caption[]{The resummed (solid) and fixed order NLO (dashed)
predictions for the light hemisphere mass distribution
$\frac{\Light}{\sigma}\frac{d\sigma}{d\Light}$ at $\mu = Q = M_Z$. 
The resummed prediction at $\mu = Q/2$ ($\mu = 2Q$) is shown as a dotted
(dot-dashed) curve. }
\label{fig:ml}
\end{figure}

\begin{figure}
\begin{center}
~\epsfig{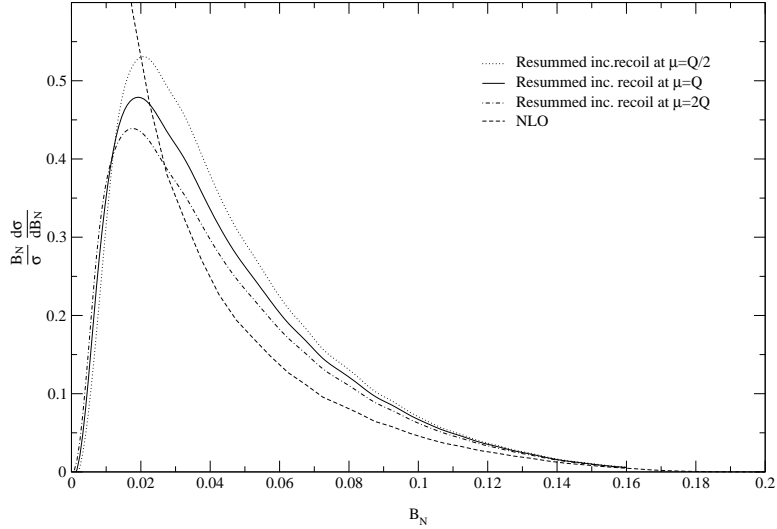}
\end{center}
\caption[]{The resummed (solid) and fixed order NLO (dashed)
predictions for the narrow jet broadening distribution
$\frac{\Bmin}{\sigma}\frac{d\sigma}{d\Bmin}$ at  $\mu = Q = M_Z$. 
The resummed prediction at $\mu = Q/2$ ($\mu = 2Q$) is shown as a dotted
(dot-dashed) curve. }
\label{fig:bn}
\end{figure}

Numerical results for the light hemisphere mass and for the narrow jet
broadening are shown in figures~\ref{fig:ml} and \ref{fig:bn} respectively.  
Throughout we set $\mu = Q = M_Z$ and use $\as(M_Z) = 0.118$ corresponding to the
current world average. 
The next-to-leading order result which diverges at small
values of the event shapes  is taken from \cite{eerad2} and is evaluated at $\mu
= Q$. 

Figures~\ref{fig:ml} and \ref{fig:bn} show that the
resummations are extremely important for $\Light < 0.01$ and $\Bmin < 0.02$. 
Rather than the divergent fixed order prediction, we have the more physical
resummed result that the probability of finding events with no radiation (very
small values of $\Light$ and $\Bmin$) are vanishingly small. For the reference
value of $\mu =Q$,  the
peak position of the $\Light$ distribution 
occurs at $\Light = 0.01$ with a height of 0.33, while for
$\Bmin$ the peak occurs at $\Bmin = 0.02$ with a value of 0.48. At larger
values, the resummation changes the NLO prediction by a more moderate amount
indicating that uncalculated higher order corrections are under control. At
very large values of $O$, the resummed and NLO predictions coincide because of
the matching procedure of Eq.~(\ref{eq:mod}). 

To illustrate the residual renormalisation scale dependence, we also show the effect
of varying $\mu$ by a factor of 2 either side of the reference value $\mu = Q$.  
We see that around the peak region, different scale choices alter the prediction
by $\pm 10\%$.  

\begin{figure} 
\begin{center}
~\epsfig{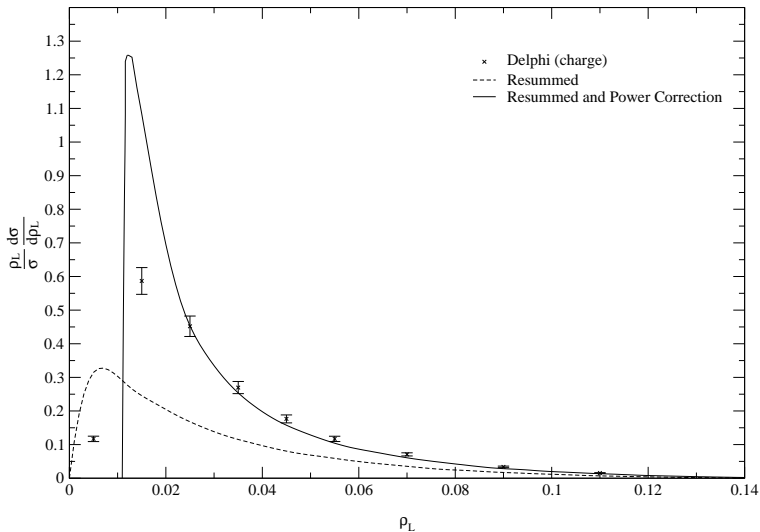}
\end{center}
\caption[]{The resummed
prediction for the light hemisphere mass distribution
$\frac{\Light}{\sigma}\frac{d\sigma}{d\Light}$ at $\mu = Q = M_Z$ modified by a
non-perturbative power correction $\Light \to \Light + 1~{\rm GeV}/Q$. The resummed prediction without
power correction is shown as a dashed line.
For comparison, we also show the 
charged hadron data collected at the $Z$ resonance by the DELPHI collaboration
\cite{delphi}.}
\label{fig:mlpc}
\end{figure}

\begin{figure} 
\begin{center}
~\epsfig{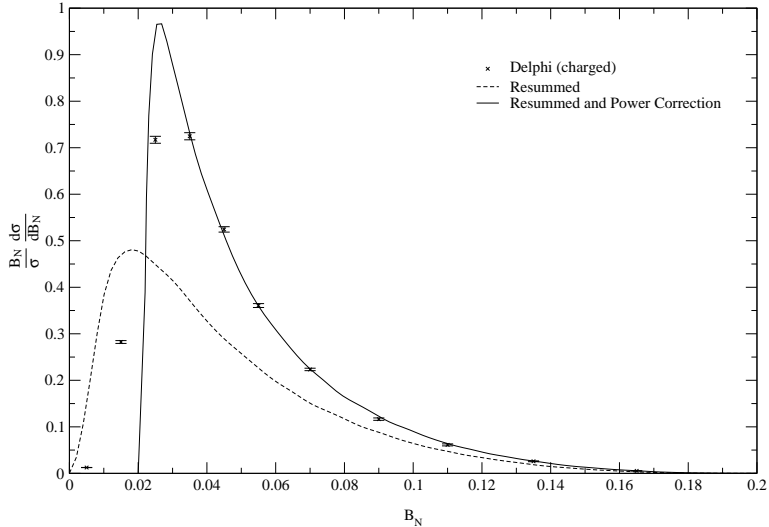}
\end{center}
\caption[]{The resummed
prediction for the narrow jet broadening distribution
$\frac{\Bmin}{\sigma}\frac{d\sigma}{d\Bmin}$ at $\mu =Q = M_Z$ modified by a
non-perturbative power correction 
$\Bmin \to \Bmin + 0.3~{\rm GeV}\ln(1/\Bmin)/Q$. The resummed prediction without
power correction is shown as a dashed line.
For comparison, we also show the 
charged hadron data collected at the $Z$ resonance by the DELPHI collaboration
\cite{delphi}.}
\label{fig:bnpc}
\end{figure}

The infrared resummation significantly improves the perturbative prediction for the
event shape observable.   However, when  comparing with experimental data we should
be aware that important non-perturbative hadronisation corrections are present. 
The effect of hadronisation on the distribution is to shift the value of the
observable away from the two-jet region,
\begin{equation} 
O \to O + O_{\rm NP}, 
\end{equation} 
where the non-perturbative correction depends on the typical hadron scale
$\O{1~{\rm GeV}}$ and is suppressed by a power of $Q$.
In
principle these power corrections  can be estimated using the dispersive
approach of Ref.~\cite{thrustpc} where a non-perturbative parameter $\mu_I$ is
introduced to describe the running of $\as$ in the infrared region. For the
associated  three jet variables the non-perturbative corrections are typically
estimated to be $\O{1~{\rm GeV}/Q}$ for $\Heavy$ \cite{thrustpc} and
$\O{0.3~{\rm GeV}\ln(1/\Bmax)/Q}$ for $\Bmax$ \cite{widepc} and arise through the
hadronisation of one of the two jets in the event.   
Because the four-jet event shapes are largely related to what happens in the
second jet, we might expect that the hadronisation corrections are similar.
To illustrate the potential
effects of hadronisation in the four-jet event shapes,  we just transfer these
corrections directly so that,
\begin{eqnarray} 
\Light &\to& \Light + \frac{1~{\rm GeV}}{Q},\\ 
\Bmin &\to& \Bmin + \frac{0.3~{\rm GeV} \ln(1/\Bmin)}{Q}. 
\end{eqnarray} 
The simplified 
hadronisation correction applied to the resummed distributions for
the light hemisphere mass and narrow jet broadening with $\mu = Q$ is shown in
figures~\ref{fig:mlpc} and \ref{fig:bnpc} respectively.   To emphasize the 
dramatic effect the power correction has, we also show the  uncorrected
predictions.
There are two effects. 
First the distribution is shifted to the right by an amount $O_{\rm NP}$ and second,
the distribution is rescaled by a factor $(O+O_{\rm NP})/O$.  In the region of the
turnover where $O$ is of the same order as $O_{\rm NP}$  there is an enhancement 
of ${\cal O}$(100\%).   The hadronisation correction is smaller at
larger values of $O$.

For comparison, we also show the charged hadron data collected by the DELPHI
Collaboration \cite{delphi} at the $Z$ resonance.   We see remarkable agreement
(strikingly so in view of the simplified hadronisation correction applied
here).    The only discrepancy is at very small values of $O < O_{\rm NP}$
where individual hadrons in the light/narrow hemisphere will
significantly affect the value of $O$.
We do not expect to successfully describe such events. 

\section{Conclusions}

In this paper, we presented predictions for the light hemisphere mass and the
narrow jet broadening distributions where the infrared logarithms have been
resummed to next-to-leading logarithmic order using the coherent
branching formalism.  The resummed expressions do not exponentiate, but involve
the difference of exponential factors so that the perturbative series starts at
$\O{\as^2}$.   The expressions presented in sections 3 and 4 resum the leading
$\O{\as^n L^{2n-1}}$ and next-to-leading $\O{\as^n
L^{2n-2}}$ infrared logarithms to all orders in the coupling constant.   Taken in conjunction with
the  fixed order $\O{\as^3}$ non-logarithmic terms, the first neglected term is
of  $\O{\as^4 L^6}$. The numerical results shown in Figs.~\ref{fig:ml} and
\ref{fig:bn}  indicate that the resummation effects are sizeable at small
values of the event shape parameter and that the resummation procedure
significantly improves the perturbative prediction.  

However, because important non-perturbative hadronisation corrections are
\linebreak present, resummation alone is insufficient to describe the
experimental data.  Applying simple power corrections similar to those
obtained for the heavy hemisphere mass and the wide jet broadening gives good
qualitative agreement with the available data from LEP. We anticipate that
the improved theoretical description of the four-jet event shape
distributions presented here can be combined with a more sophisticated
hadronisation correction based on the dispersive approach of
Ref.~\cite{thrustpc}.  The data from LEP can then be used to further test the
structure of QCD in four-jet like events and extract values of the strong
coupling constant.

\section*{Acknowledgements}

We thank Mike Seymour and David Summers for useful and stimulating discussions.  
SJB thanks the UK Particle Physics and Astronomy  Research Council for research
studentships.   This work was supported in part by the EU Fourth Framework 
Programme `Training and Mobility of Researchers',  Network `Quantum Chromodynamics
and the Deep Structure of  Elementary Particles', contract FMRX-CT98-0194 
(DG-12-MIHT).

\end{document}